\documentclass[twocolumn,aps,nofootinbib,superscriptaddress,tightenlines,notitlepage,pra,longbibliography]{revtex4-2}

\usepackage[english]{babel}
\usepackage{amsmath} 				

\usepackage{amssymb} 				
\usepackage{amsthm}
\usepackage{array}       				
\usepackage{graphicx}  					
\usepackage{xcolor}       			
\usepackage[utf8]{inputenc} 			
\usepackage{url}     
\usepackage{times}
\usepackage{stmaryrd}

\usepackage{svg}

\usepackage[normalem]{ulem}

\usepackage[colorlinks=true,linkcolor=teal,citecolor=teal,urlcolor=teal]{hyperref}
\usepackage{bbold}
\usepackage{bm}
\usepackage{bbm}
\usepackage{microtype}

\usepackage[shortlabels]{enumitem}
\usepackage{braket}
\usepackage{mathtools}

\usepackage{tikz}
\usetikzlibrary{positioning}

\usepackage{hyperref}

\newtheorem{theorem}{Theorem}

\newcommand{\ketbra}[2]{\ket{#1}\!\!\bra{#2}}
\newcommand{\abs}[1]{\lvert #1 \rvert}
\newcommand{\norm}[1]{\lVert #1 \rVert}

\newcommand{\trace}[1]{\mathrm{tr}\!\left(#1\right)}
\newcommand{\partialtrace}[2]{\mathrm{tr}_{#1}\!\left(#2\right)}

\newcommand{\ee}{\mathrm{e}}
\newcommand{\ii}{\mathrm{i}}

\definecolor{richard}{RGB}{164,12,52}

\begin{document}

\title{In the shadow of the Hadamard test:\\Using the garbage state for good and further modifications}

    \author{\href{https://orcid.org/0000-0002-8706-1732}{Paul\ K.\ Faehrmann}}
	\affiliation{Dahlem Center for Complex Quantum Systems, Freie Universität Berlin, 14195 Berlin, Germany}
    \affiliation{Institute for Integrated Circuits and Quantum Computing, Johannes Kepler University Linz, Austria}

    \author{\href{https://orcid.org/0000-0003-3033-1292}{Jens\ Eisert}}
	\affiliation{Dahlem Center for Complex Quantum Systems, Freie Universität Berlin, 14195 Berlin, Germany}
	\affiliation{Helmholtz-Zentrum Berlin f{\"u}r Materialien und Energie, Hahn-Meitner-Platz 1, 14109 Berlin, Germany}

    \author{\href{https://orcid.org/0000-0002-8291-648X}{Richard\ Kueng}}
    \affiliation{Institute for Integrated Circuits and Quantum Computing, Johannes Kepler University Linz, Austria}

\begin{abstract}
The Hadamard test is naturally suited for the intermediate regime between the current era of noisy quantum devices and complete fault tolerance. Its applications use measurements of the auxiliary qubit to extract information, but disregard the system register completely. Separate advances in classical representations of quantum states via classical shadows allow the implementation of even global classical shadows with shallow circuits. This work combines the Hadamard test on a single auxiliary readout qubit with classical shadows on the remaining $n$-qubit work register. We argue that this combination inherits the best of both worlds and discuss statistical phase estimation as a vignette application. There, we can use the Hadamard test to estimate eigenvalues on the auxiliary qubit, while classical shadows on the remaining $n$ qubits provide access to additional features such as, (i) fidelity with certain pure quantum states, (ii) the initial state's energy and (iii) how pure and how close the initial state is to an eigenstate of the Hamiltonian. Finally, we also discuss how anti-controlled unitaries can further augment this framework. 
\end{abstract}

\maketitle

\section{Introduction}
After the recent demonstration of the first logical quantum computations~\cite{reichardtDemonstrationQuantumComputation2024,bluvsteinLogicalQuantumProcessor2024,acharyaQuantumErrorCorrection2024,reichardtLogicalComputationDemonstrated2024,puttermanHardwareefficientQuantumError2025}, we are on the verge of leaving the era of noisy, intermediate-scale quantum devices (NISQ)~\cite{preskillQuantumComputingNISQ2018b} and entering the era of early fault tolerance or \emph{intermediate scale-quantum devices} (ISQ)~\cite{FromNISQtoISQs} and the \emph{megaquop machine}~\cite{preskillNISQMegaquopMachine2025}, a quantum device that can perform of the order of a million of quantum operations.
With only a few error-corrected qubits but intermediate-sized quantum devices available, a natural next step is to combine noisy and error-corrected registers to implement more and more intricate quantum algorithms~\cite{koukoulekidisFrameworkPartialError2023}.
Thus, the question arises which quantum algorithms best suit these architectures.

Arguably, the Hadamard test and variations thereof are ideal candidates because they use a single auxiliary qubit that gets entangled with all other qubits. But only this qubit is measured in the end. 
Its ability to be used as a subroutine in algorithms that classically combine measurements to reconstruct expectation values of linear combinations of unitaries has sparked the development of several quantum algorithms aimed at resource-efficient energy estimation~\cite{kitaevQuantumMeasurementsAbelian1995,svoreFasterPhaseEstimation2014,wiebeEfficientBayesianPhase2016,linHeisenbergLimitedGroundStateEnergy2022,wanRandomizedQuantumAlgorithm2022,clintonPhaseEstimationLocal2023,wangQuantumAlgorithmGround2023,guntherPhaseEstimationPartially2025}, sampling from matrix functions such as for solving linear systems~\cite{wangQubitEfficientRandomizedQuantum2024}, quantum dynamics~\cite{faehrmannRandomizingMultiproductFormulas2022}, Gibbs state preparation or properties thereof~\cite{chowdhuryVariationalQuantumAlgorithm2020,wangVariationalQuantumGibbs2021,consiglioVariationalQuantumAlgorithms2023,wangQubitEfficientRandomizedQuantum2024}, estimating dynamical correlations via Green's functions~\cite{bauerHybridQuantumClassicalApproach2016,wangQubitEfficientRandomizedQuantum2024},  linear response of quantum systems~\cite{baroniNuclearTwoPoint2022}, computing the density of states~\cite{gohDirectEstimationDensity2024}, entanglement spectroscopy and the estimation of $\alpha$-Renyi entropies~\cite{subasiEntanglementSpectroscopyDepthtwo2019,subramanianQuantumAlgorithmEstimating2021a}, approximating the Jones polynomial~\cite{aharonovPolynomialQuantumAlgorithm2006,shorEstimatingJonesPolynomials2008}, as well as giving rise to the one clean qubit model of quantum computation (DQC1)~\cite{knillPowerOneBit1998}.

Furthermore, allowing for additional auxiliary qubits and circuit depth, the Hadamard test can be extended to make use of the existing resources both in the intermediate fault-tolerant regime~\cite{wangAcceleratedVariationalQuantum2019} or in the far-term application of the original quantum phase estimation algorithm~\cite{Nielsen_Chuang_2010}, or even simplified using phase retrieval techniques to avoid the controlled unitary~\cite{clintonQuantumPhaseEstimation2024}.
These algorithms mostly use adaptations of the Hadamard test by measuring the auxiliary qubit to estimate expectation values of the form $\trace{\ee^{\ii Ht_i}\rho}$ for some Hamiltonian $H$ and state $\rho$ with different evolution times $t_i$ to construct Fourier approximations $F(H,t)$ of desired spectral functions in classical post-processing.
However, only information from the auxiliary qubit is extracted for this trace estimation since the garbled post-measurement state of the system register is of no apparent use.
Notable exceptions are applications using system register measurements for error mitigation within verified phase estimation~\cite{pollaOptimizingInformationExtracted2023,obrienErrorMitigationVerified2021} and the combination of the generalized swap test with shadows on the copies of the input state~\cite{zhouHybridFrameworkEstimating2024}.
\begin{figure*}[t!]
    \centering
    \includegraphics[width=.8\linewidth]{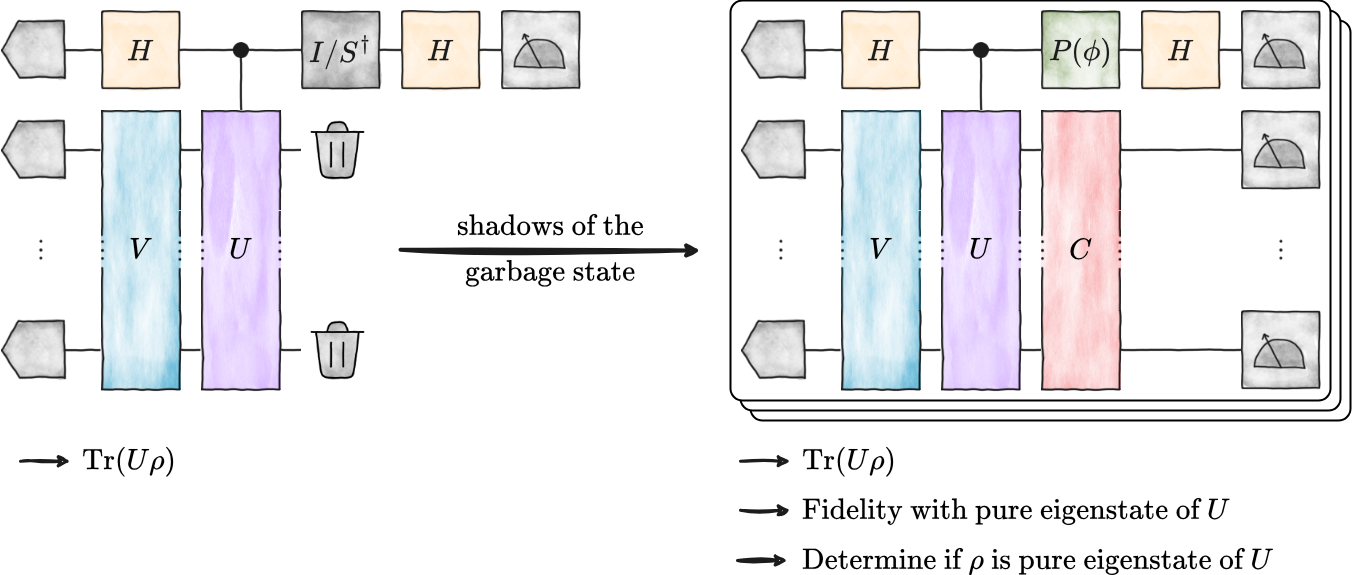}
    \caption{Cartoon illustration of the proposed adaptation of the Hadamard test: \emph{(left)} The standard Hadamard test circuit allows for the estimation of $\mathrm{Re}(\trace{U\rho})$ or $\mathrm{Im}(\trace{U\rho})$ when using $\phi=0$ and $\phi=\pi/2$ respectively. 
    Here, $V$ labels the state preparation unitary, but the state can also be mixed.
    \emph{(right)} Instead of disregarding the system register, we can perform local or global shadow estimation of the post-measurement state by applying random local or global Clifford gates ($C$) and thereby extract so-far unused information.}
    \label{fig:had_test}
\end{figure*}

At the same time, the advent of classical shadows as a tool to efficiently construct approximate classical descriptions of quantum states using very few measurements has impressively showcased the fundamental power of quantum measurements in conjunction with classical post-processing \cite{huangPredictingManyProperties2020c,elbenRandomizedMeasurementToolbox2023,morrisSelectiveQuantumState2020,painiEstimatingExpectationValues2021}, leading to hybrid quantum-classical algorithms enhancing quantum devices with classical shadows~\cite{sackAvoidingBarrenPlateaus2022b,chanAlgorithmicShadowSpectroscopy2024,faehrmannShorttimeSimulationQuantum2024,zhouHybridFrameworkEstimating2024}.
After the recent breakthrough of Ref.~\cite{schusterRandomUnitariesExtremely2024a} and previous shallow constructions~\cite{bertoniShallowShadowsExpectation2024a}, even global classical shadows are provably accessible with extremely low-depth quantum circuits.

Equipped with these tools, it is time to revisit previous quantum algorithms and search for gems in so far unmeasured output states.
We summarize the Hadamard test and its applications in Section~\ref{sec:had_intro}.
In Section~\ref{sec:shadows}, we then show that the standard presentation of the Hadamard test can be expanded by including classical shadows of the system register. 
In classical post-processing, combining the auxiliary register's measurement results with the system register's shadow measurement allows for estimating additional quantities, inheriting the sample complexity bounds of classical shadows.

To demonstrate the usefulness of this additional information, we sketch three applications: (1) simultaneous estimation of $\trace{F(H,t)\rho}$ for a Fourier approximation $F(H,t)$ of some spectral function and fidelities with known eigenstates (using global shadows), where energy estimation following Refs.~\cite{linHeisenbergLimitedGroundStateEnergy2022,wanRandomizedQuantumAlgorithm2022,wangQuantumAlgorithmGround2023} is a specific example, (2) combining $\trace{F(H,t)\rho}$ with an estimation of $\trace{H\rho}$ (using local shadows), and (3) additionally estimating a measure of purity and eigenstateness (using local shadows).

Finally, we discuss another modification of the standard Hadamard test used in Refs.~\cite{faehrmannRandomizingMultiproductFormulas2022,wangQubitEfficientRandomizedQuantum2024,chakrabortyImplementingAnyLinear2024} and how applying anti-controlled unitaries can lead to further applications in comparing spectra of unitaries and determining eigenstateness of the input state in Section~\ref{sec:anti-control}.

\section{The Hadamard test and its uses}
\label{sec:had_intro}
To perform the standard Hadamard test shown in Figure~\ref{fig:had_test} \emph{(left)}, we repeat the quantum circuit with $\phi=0$ (i.e.\ $P(\phi)=P(0)=\mathbb{I}$) and estimate the outcome probabilities of the measurement of the auxiliary qubit which are given by
\begin{align}
    p(0)&=\frac{1}{2}\left(\trace{\rho}+\mathrm{Re}(\trace{U \rho})\right),\\
    p(1)&= \frac{1}{2}\left(\trace{\rho}-\mathrm{Re}(\trace{U \rho})\right),
\end{align}
and then obtain $\mathrm{Re}(\trace{U\rho})=p(0)-p(1)$  in classical post-processing by effectively estimating the Pauli-$Z$ expectation value of the auxiliary qubit while tracing over the system register.
Repeating the procedure for $\phi=\pi/2$ allows for the estimation of $\mathrm{Im}(\trace{U\rho})$.

The linearity in the unitary $U$ is precisely the reason why the Hadamard test has led to so many algorithms aimed at the resource-efficient use of intermediate-scale quantum devices.
Given a \emph{linear combination of unitaries} (LCU)  $M_\mathrm{LCU}=\sum\alpha_iU_i$, repetitions of the Hadamard test for the individual $ U_i$'s allow for an estimation of $\trace{M_\mathrm{LCU}\rho}$ in classical post-processing.
This bypasses the block-encoding and amplitude amplification procedures required for a coherent application of $M_\mathrm{LCU}$.
Furthermore, the linearity also allows for importance sampling procedures not only from $M_\mathrm{LCU}$ but also of the $ U_i$'s themselves.  
This can help reduce circuit depths and lead to novel algorithms, as demonstrated in Ref.~\cite{wanRandomizedQuantumAlgorithm2022}.

Whereas quantum signal processing and qubitization approaches focus on Chebychev polynomials as a basis to construct linear combinations of unitaries to approximate spectral functions~\cite{lowHamiltonianSimulationQubitization2019b,lowOptimalHamiltonianSimulation2017b,martynGrandUnificationQuantum2021b,gilyenQuantumSingularValue2019a}, the Hadamard test is especially suited to implement Fourier approximations thereof. 
As such, most proposed applications use $U_i=\ee^{\ii Ht_i}$ to construct useful Fourier approximations, such as those of step-functions or filter functions that allow projections into (potentially low-energy) subspaces or for eigenvalue thresholding~\cite{linHeisenbergLimitedGroundStateEnergy2022,wanRandomizedQuantumAlgorithm2022,wangQuantumAlgorithmGround2023}.
These applications thus combine the Hadamard test with Hamiltonian simulation to access a different basis in which to approximate spectral functions in classical post-processing.

The construction of further interesting Fourier functions in a highly relevant and ongoing research topic and any findings in this direction fit the following discussion of adding system register measurements to the Hadamard test.

\section{The Hadamard test in the light of shadows}
\label{sec:shadows}
While the standard application of the Hadamard test ends with the computation of $\trace{U\rho}$, the adaptations of Refs.~\cite{faehrmannRandomizingMultiproductFormulas2022,wangQubitEfficientRandomizedQuantum2024} have shown that appropriate measurements of the system register can lead to the estimation of so far unavailable quantities, making better use of the output state of the Hadamard circuit before measurement, given by
\begin{align}
\label{eq:had-output}
    {\rho_\mathrm{out}} = \frac{1}{4}\biggl(&\ketbra{0}{0}\otimes (I + U\ee^{\ii\phi})\rho(I + U\ee^{\ii\phi})^\dagger\\ \nonumber
    +&\ketbra{0}{1}\otimes (I + U\ee^{\ii\phi})\rho(I - U\ee^{\ii\phi})^\dagger\\ \nonumber
    +&\ketbra{1}{0}\otimes (I - U\ee^{\ii\phi})\rho(I + U\ee^{\ii\phi})^\dagger\\ \nonumber
    +&\ketbra{1}{1}\otimes (I - U\ee^{\ii\phi})\rho(I - U\ee^{\ii\phi})^\dagger \biggr).
\end{align}
Thus, we would like to consider the additional possibilities the Hadamard test circuit offers, including estimating different Pauli expectation values on the auxiliary qubit, tracing out the auxiliary register, and adding observable measurements (that can also result in classical shadows) to the system register.

While a thorough, step-by-step discussion of all of these cases is presented in Appendices~\ref{app:vanilla_had} to \ref{app:system_register} and a list summarizing all available quantities shown in Appendix~\ref{app:summary}, we would like to focus on two specific applications using the post-measurement (potentially non-normalized) states
\begin{align}
   \rho (Z) &=\partialtrace{\mathrm{aux}}{Z_\mathrm{aux}\rho_\mathrm{out}}= \frac{1}{2} \left(   U\rho \ee^{\ii\phi} + \rho U^\dagger \ee^{-\ii\phi}  \right)\label{eq:rhoZ},\\
   \rho (I) &=\partialtrace{\mathrm{aux}}{\rho_\mathrm{out}}= \frac{1}{2} \left( \rho+U\rho U^\dagger\right)\label{eq:rhoI},
\end{align}
where $\rho_\mathrm{out}$ denotes the output state of the Hadamard test before any measurements.

Although we could instead perform several Hadamard tests in parallel and physically linearly combine their post-measurement states depending on the measurement outcome, classical shadows allow a direct use of these post-measurement states, one that is available for all of the algorithms mentioned above, at the (negligible) cost of also measuring the system register.
Furthermore, the well-established literature on classical shadows provides rigorous error bounds and guarantees that are directly applicable here as well and are summarized in Appendix~\ref{app:shadow_complexity}.

\subsection{Using system register measurements for fidelity estimation}
To showcase the use of obtaining classical shadows of such post-measurement states, we turn toward a concrete example: combining the estimation of expectation values of linear combinations of unitaries with fidelities with pure eigenstates of $H$.

We begin by first discussing the setting of using only a single, fixed $U$ within the Hadamard test.
Combining the measurement of the auxiliary qubit of the output state $\rho_\mathrm{out}$ in the Pauli-$Z$ basis (see Eq.~\eqref{eq:Z_measurement}) results in a statistical estimator for
\begin{equation}
    \braket{Z \otimes I^{\otimes n}}_{\rho_{\mathrm{out}}}
=\trace{\rho (Z)} = \mathrm{Re} \left( \trace{ \ee^{\ii \phi} U \rho } \right)
\end{equation}
and consequently in a statistical estimator of $\mathrm{Im} \left( \trace{ U \rho } \right)$ for $\phi=\pi/2$ and thus an effective Pauli-$Y$ measurement, as before.

However, we can also measure some observable $O$ on the remaining $n$-qubit system register, resulting in 
\begin{equation}
    \braket{Z \otimes O}_{\rho_{\mathrm{out}}}
=\trace{O\rho (Z)} = \mathrm{Re} \left( \trace{ \ee^{\ii \phi}O U \rho }\right).
\end{equation}
Note that a measurement of $O$ still allows for disregarding that measurement information, thus effectively tracing over the system register, and therefore does not affect the original goal of estimating $\trace{U \rho}$.

Now, just as we use separate estimations of the real and imaginary part of $\trace{U \rho}$ for its estimation in classical post-processing, we can combine the system register measurements to obtain $\trace{OU\rho}$ and consequently classical shadows of $\mathrm{Re}(U)\rho$, $\mathrm{Im}(U)\rho$ and $U\rho$.

Since this quantity is still linear in $U$, we can again extend these results to linear combinations of different unitaries, thereby obtaining the same for $M_\mathrm{LCU}=\sum\alpha_iU_i$.
Thus, when randomly choosing which $U_i$ of the linear combination to implement and further randomizing the observable $O$, we can obtain information about the (non-normalized) state $\widetilde{\rho}=M_\mathrm{LCU}\rho$.

Let us make this more concrete for $U_i=\ee^{\ii H t_i}$ by restricting to rank-one observables $O = \ketbra{\lambda}{\lambda}$ to obtain
\begin{align}
    \braket{Z \otimes \ketbra{\lambda}{\lambda}}_{{\rho_\mathrm{out},i}}
&= \trace{\ketbra{\lambda}{\lambda} \rho_i (Z)}\\
&=\mathrm{Re}\left(\trace{\ketbra{\lambda}{\lambda}\ee^{\ii (Ht_i+I\phi)}\rho}\right).
\nonumber
\end{align}
Besides restricting to a single $\ket{\lambda}$, which may or may not be preparable with a shallow circuit, we can also use global shadows on the system register, which are now available with shallow circuits thanks to Ref.~\cite{schusterRandomUnitariesExtremely2024a} or Ref.~\cite{bertoniShallowShadowsExpectation2024a} to estimate this quantity for arbitrary product states. 

Now, if we assume that $\ket{\lambda}$ is a pure eigenstate of $H$ with known eigenvalue $E$, we find
\begin{align}
\nonumber
    \braket{Z \otimes \ketbra{\lambda}{\lambda}}_{{\rho_\mathrm{out},i}}
&=\mathrm{Re}\left(\ee^{\ii (Et_i+\phi)}\bra{\lambda}\rho\ket{\lambda}\right)\\
&=\cos(Et_i+\phi)F(\rho,\ketbra{\lambda}{\lambda})\label{eq:fidelity_est},
\end{align}
which, since we know the factor $\cos(Et_i+\phi)$, allows us to estimate the fidelity $F$ between the initial state $\rho$ and known, pure eigenstate $\ket{\lambda}$.
Consequently, we can obtain local (energy) and global (fidelity) simultaneously using the Hadamard test.
More generally, when $M_\mathrm{LCU}=\sum{\alpha_i}\ee^{\ii H t_i}$, we can obtain $\mathrm{Re}(M_\mathrm{LCU}(\lambda))F(\rho,\ketbra{\lambda}{\lambda})$, $\mathrm{Im}(M_\mathrm{LCU}(\lambda))F(\rho,\ketbra{\lambda}{\lambda})$ and linear combinations thereof.

Thus, when $M_\mathrm{LCU}$ is a Fourier series of a threshold function projecting the initial state $\rho$ into a (potentially low-energy) subspace, fidelities with pure eigenstates of $H$ with known energy within the same subspace can be estimated.
However, since the threshold function yields one for all eigenstates within the selected subspace, we do not need to know the energy explicitly but only that it is within the subspace.

The only additional cost of obtaining these additional estimates is the cost of estimating the required observables on the system register, which, due to recent shallow constructions for global shadows, does not constitute a bottleneck in practice.

In general, the same procedure can be used for extended Hadamard test circuits with more than a single auxiliary qubit and controlled unitary. However, it is essential to note that the number of measurement outcomes scales exponentially, decreasing the resolution for each post-measurement shadow.
The advantage remains that only a single quantum circuit is required to extract both the original information of these circuits and the information tractable with linear combinations of the post-measurement shadows.

It is further important to note that stochastic phase estimation achieves Heisenberg limit scaling and thus requires only $\mathcal{O}(\epsilon^{-1})$ samples to obtain phase knowledge with an error of at most $\epsilon$~\cite{linHeisenbergLimitedGroundStateEnergy2022,wangQuantumAlgorithmGround2023}. In contrast, classical shadows or even simple observable estimation by measurement are sampling procedures, requiring $\mathcal{O}(\epsilon^{-2})$ samples to achieve the same error guarantees.
Thus, when focusing on stochastic phase estimation, the additional quantities accessible via system register measurements can only be estimated up to an error of $\mathcal{O}(\sqrt{\epsilon})$.

\subsection{Using system register measurements for energy estimation}
So far, we have used global shadows of the post-measurement state $\rho(Z)$ to extract information about fidelities. 
However, as discussed in the appendices, we can also trace out the auxiliary register to obtain
\begin{equation}
\rho (I) =\partialtrace{\mathrm{aux}}{\rho_\mathrm{out}}= \frac{1}{2} \left( \rho+U\rho U^\dagger\right).
\end{equation}
For local Hamiltonians and linear combinations of $U_i=\ee^{\ii H t_i}$, local classical shadows, or any other measurement scheme for energy estimation, of the system register can then be used to obtain an additional estimate of the state's energy, since for $O=H$, obtainable in classical post-processing due to the linearity in $O$, we have
\begin{equation*}
    \trace{H\rho_i(I)}=\frac{1}{2}\left(\trace{H\rho}+\trace{H\ee^{\ii H t_i}\rho \ee^{-\ii Ht_i}}\right)=\trace{H\rho}.
\end{equation*}
Consequently, as a byproduct of estimating $\trace{M_\mathrm{LCU}\rho}$, we can obtain an energy estimate of $\rho$, requiring only additional random Pauli measurements of the $n$-qubit work register.
Similarly, the expectation value of other operators commuting with the Hamiltonian can be estimated.

It is important to stress that one of the main early applications of the Hadamard test is stochastic phase estimation~\cite{linHeisenbergLimitedGroundStateEnergy2022,wangQuantumAlgorithmGround2023}, whose goal is the estimation of a state's eigenenergies and, more importantly, ground state energies. 
As such, these applications already generate energy information with a Heisenberg scaling of $\mathcal{O}(\epsilon^{-1})$ samples. In contrast, the above use of shadows requires $\mathcal{O}(\epsilon^{-1})$ samples to achieve the same accuracy. 
However, we envision using statistical phase estimation for ground state energy estimation with simultaneous estimation of the guiding state's energy $\trace{H\rho}$.
Therefore, we view this application as an add-on to more intricate phase estimation procedures and an enhancement of Hadamard circuit applications, whose purpose is not foremost in energy estimation.

\subsection{Using system register measurements for purity and eigenstateness}
Less practical but still conceptually interesting, we can further use exponentially many local shadows of $\rho (I)$ to estimate its purity~\cite{sackAvoidingBarrenPlateaus2022b}, given by
\begin{equation}
    \trace{\rho (I)^2}=\frac{1}{2}\left(\trace{\rho^2}+\trace{\rho U \rho U^\dagger}\right),
\end{equation}
which is one if and only if $\rho$ is a pure eigenstate of $U$.
Tracing out the auxiliary system thus adds these two quantum states, allowing for another quantity to be estimated, albeit not efficiently in this case, since purity estimation is known to require exponentially many copies of the state~\cite{chenExponentialSeparation}.

Since the system register can be traced over even when measured, we can again combine the estimation of this quantity with the estimation of energy.
Furthermore, even though this quantity is no longer linear in $U$, if $U_i=\ee^{\ii Ht_i}$, the simulation time does not impact whether $\rho$ is an eigenstate of $U_i$.
Since the introduced phases of $\ee^{\pm\ii \lambda t_i}$ for an eigenstate $\rho=\ketbra{\lambda}{\lambda}$ with energy $\lambda$ cancel, we can again use measurement outcomes for different $t_i$ to estimate the same quantity.

\section{Adding anti-controlled unitaries to the Hadamard test}
\label{sec:anti-control}
\begin{figure}[t!]
    \centering
    \includegraphics[width=\linewidth]{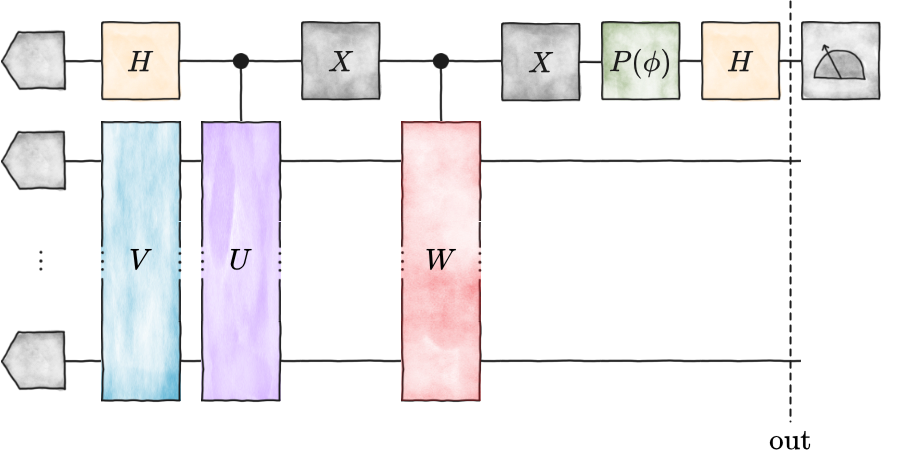}
    \caption{The Hadamard test with an additional anti-controlled unitary $W$ can help in quantum dynamics and linear algebra by linearizing the output to allow for randomized approaches~\cite{faehrmannRandomizingMultiproductFormulas2022,wangQubitEfficientRandomizedQuantum2024}, or to compare spectra of unitaries and determine eigenstateness of a state as discussed in the main text.}
    \label{fig:had_two_controls}
\end{figure}

Stepping away from system register measurements, we want to discuss another component of the Hadamard test that has not been thoroughly explored: the addition of an anti-controlled unitary $W$ (applied when the auxiliary qubit is zero instead of one), as shown in Figure~\ref{fig:had_two_controls}.
As discussed in Refs.~\cite{faehrmannRandomizingMultiproductFormulas2022,wangQubitEfficientRandomizedQuantum2024,chakrabortyImplementingAnyLinear2024}, this helps to linearize the problem of applying two linear combinations of unitaries in a randomized fashion as required for quantum dynamics or linear algebra applications.

In essence, this changes the output state of the Hadamard test from Eq.~\eqref{eq:had-output} to
\begin{align}
    {\rho_\mathrm{out}} = \frac{1}{4}\biggl(&\ketbra{0}{0}\otimes (W + U\ee^{\ii\phi})\rho(W + U\ee^{\ii\phi})^\dagger\\ \nonumber
    +&\ketbra{0}{1}\otimes (W + U\ee^{\ii\phi})\rho(W - U\ee^{\ii\phi})^\dagger\\ \nonumber
    +&\ketbra{1}{0}\otimes (W - U\ee^{\ii\phi})\rho(W + U\ee^{\ii\phi})^\dagger\\ \nonumber
    +&\ketbra{1}{1}\otimes (W - U\ee^{\ii\phi})\rho(W - U\ee^{\ii\phi})^\dagger \biggr),
\end{align}
effectively replacing the $(I\pm U\ee^{\ii\phi})$ of the unmodified Hadamard test with $(W\pm U\ee^{\ii\phi})$, which leads to additional obtainable post-measurement states and observables, summarized in Appendix~\ref{app:summary_extended}.
Besides the applications of Refs.~\cite{faehrmannRandomizingMultiproductFormulas2022,wangQubitEfficientRandomizedQuantum2024}, we envision another use in comparing spectra of different unitaries, further strengthening the point that small changes to existing algorithms can yield interesting new outcomes.

As an example, consider $U=\ee^{\ii Ht_1}$ and $W=\ee^{\ii Ht_2}$ and $\phi=0$.
Then, executing the modified Hadamard test of Fig.~\ref{fig:had_two_controls} yields
\begin{align}
    \rho(Z)&=\frac{1}{2}\left(U\rho W^\dagger + W\rho U^\dagger\right)\\
    &=\frac{1}{2}\left(\ee^{\ii Ht_1}\rho \ee^{-\ii Ht_2}+\ee^{\ii Ht_2}\rho \ee^{-\ii Ht_1}\right).
    \nonumber
\end{align}
Now, if $\rho=\ketbra{\psi}{\psi}$ with energy $\bra{\psi}H\ket{\psi}=E$,
\begin{equation}
\label{eq:Z_for_eigenstate}
\trace{\rho(Z)} = \mathrm{Re}(\ee^{\ii E (t_1-t_2)})=\cos(E(t_1-t_2)).
\end{equation}
However, if $\rho$ is not an eigenstate of $H$, e.g., when $\rho=\ketbra{\psi}{\psi}$ with
\begin{equation}
    \ket\psi=\alpha\ket{\mu}+\beta\ket{\nu}
\end{equation}
for two different eigenstates $\ket{\mu}$ and $\ket{\nu}$ with energies $E_1$ and $E_2\neq E_1$, we obtain
\begin{align}
    \trace{\rho(Z)} &= \mathrm{Re}\left(\trace{\ee^{\ii H t_1}\rho \ee^{-\ii H t_2}}\right)\\
    &=\alpha^2\cos(E_1(t_1-t_2))\!+\!\beta^2\cos(E_2(t_1-t_2)),
    \nonumber
\end{align}
since $\trace{\ketbra{\mu}{\nu}}=\braket{\mu|\nu}=0$. 
Consequently, such a setup can be used to distinguish eigenstates from non-eigenstates, and the further away the input state is from an eigenstate, i.e., the more eigenstates it can be decomposed into with non-negligible weight, the further away the measurement statistics are from the single cosine of Eq.~\eqref{eq:Z_for_eigenstate}.

\section{Discussion and outlook}

In this work, we have revisited the Hadamard test with recently developed algorithmic subroutines in mind to show that classical shadows of the system register can extract so far unused but actually highly informative information: Accepting that accessible information has been left ignored so far that can be exploited to improve schemes adds an exciting new twist to the Hadamard test.
Furthermore, we have discussed how slight modifications of this well-known circuit can lead to many new applications.
Especially for early fault-tolerant quantum devices with only a few error-corrected (and thus few auxiliary qubits) and sampling-based quantum algorithms, this approach allows for the extraction of additional information without additional quantum circuits or 
the enhancement of previous applications using only auxiliary qubit measurements. 
Additionally, we have revisited modifying the Hadamard test to contain anti-controlled unitaries to show that this slight adaptation also yields new applications. 
These insights further motivate revisiting well-known algorithms and subroutines and exploring slight modifications thereof.

These results further demonstrate the 
power of quantum measurement and motivate the continued study of quantum algorithms that disregard qubit registers without measurements. 
This mindset opens up novel possibilities, especially using recent breakthroughs enabling both global and local classical shadows with shallow circuits. 
We hope the tools and ideas presented here will contribute to bringing NISQ devices into the next-level regime.
\newline

\paragraph*{Acknowledgments.}
We thank Stefano Polla for insightful discussions at QCTiP 2025. 
The Berlin team has been supported by the BMWK (PlanQK), the DFG (CRC 183), the BMBF (QPIC-1, MuniQC-Atoms, HYBRID), the Quantum Flagship (Millenion, Pasquans2), QuantERA (HQCC), and the ERC (DebuQC). RiK has been supported by the Austrian Science fund via SFB BeyondC (Grant-DOI 10.55776/F71), the projects QuantumReady (FFG 896217) and High-Performance integrated Quantum Computing (FFG 897481) within Quantum Austria (both managed by the FFG), as well as the ERC (q-shadows).

%

\appendix

\section{The Hadamard test as known from the literature}
\label{app:vanilla_had}
\begin{figure}[t!]
    \centering
    \includegraphics[width=0.8\linewidth]{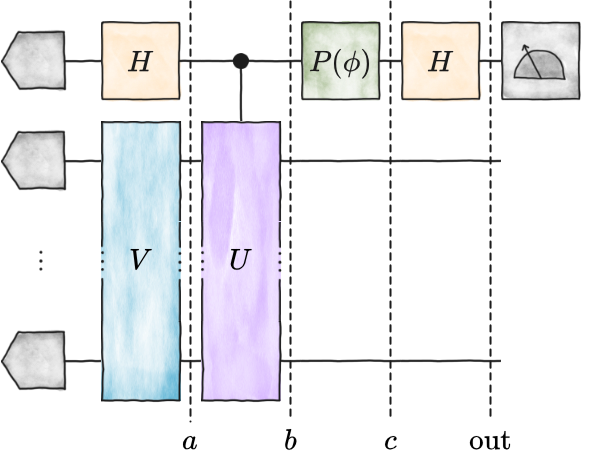}
    \caption{The standard Hadamard test.}
    \label{fig:vanilla_hadamard}
\end{figure}
For completeness, let us go through the step-by-step derivation of the Hadamard test.
As shown in Figure~\ref{fig:vanilla_hadamard}, given that $V$ prepares the initial possibly mixed state of the system $\rho$, we begin with
\begin{align}
    \rho_a = \frac{1}{2}\biggl(&\ketbra{0}{0}\otimes \rho 
    +\ketbra{0}{1}\otimes \rho \\ \nonumber
    +&\ketbra{1}{0}\otimes  \rho 
    +\ketbra{1}{1}\otimes \rho \biggr),
\end{align}
on which the controlled application of $U$ acts by adding a $U$ or $U^\dagger$ for each $\ket{1}$ and $\bra{1}$ of the auxiliary qubit respectively, i.e.,
\begin{align}
    \rho_b = \frac{1}{2}\biggl(&\ketbra{0}{0}\otimes \rho 
    +\ketbra{0}{1}\otimes \rho U^\dagger \\ \nonumber
    +&\ketbra{1}{0}\otimes  U\rho 
    +\ketbra{1}{1}\otimes U\rho U^\dagger \biggr).
\end{align}
The phase shift gate $P(\phi)$ maps 
\begin{eqnarray}
\ket{0}&\mapsto&\ket{0},\\
\ket{1}&\mapsto& e^{\ii \phi}\ket{1},
\end{eqnarray}
which, applied to the auxiliary qubit, results in
\begin{align}
    \rho_c = \frac{1}{2}\biggl(&\ketbra{0}{0}\otimes \rho 
    +\ketbra{0}{1}\otimes \rho U^\dagger \ee^{-\ii\phi}\\ \nonumber
    +&\ketbra{1}{0}\otimes  U\rho \ee^{\ii\phi}
    +\ketbra{1}{1}\otimes U\rho U^\dagger \biggr).
\end{align}
The application of another Hadamard gate then changes the basis of the auxiliary qubit to
\begin{align}
    \label{eq:had-output-app}
    {\rho_\mathrm{out}} = \frac{1}{4}\biggl(&\ketbra{0}{0}\otimes (I + U\ee^{\ii\phi})\rho(I + U\ee^{\ii\phi})^\dagger\\ \nonumber
    +&\ketbra{0}{1}\otimes (I + U\ee^{\ii\phi})\rho(I - U\ee^{\ii\phi})^\dagger\\ \nonumber
    +&\ketbra{1}{0}\otimes (I - U\ee^{\ii\phi})\rho(I + U\ee^{\ii\phi})^\dagger\\ \nonumber
    +&\ketbra{1}{1}\otimes (I - U\ee^{\ii\phi})\rho(I - U\ee^{\ii\phi})^\dagger \biggr)\\
    = \frac{1}{4}\biggl(&\ketbra{0}{0}\otimes \left(\rho  + U\rho \ee^{\ii\phi} + \rho U^\dagger \ee^{-\ii\phi} +U\rho U^\dagger\right)
    \nonumber
    \\ \nonumber
    +&\ketbra{0}{1}\otimes \left(\rho  + U\rho \ee^{\ii\phi} - \rho U^\dagger \ee^{-\ii\phi}-U\rho U^\dagger\right)\\ \nonumber
    +&\ketbra{1}{0}\otimes \left(\rho - U\rho \ee^{\ii\phi}+ \rho U^\dagger \ee^{-\ii\phi} -U\rho U^\dagger\right)\\ \nonumber
    +&\ketbra{1}{1}\otimes \left(\rho - U\rho \ee^{\ii\phi}-  \rho U^\dagger \ee^{-\ii\phi} +U\rho U^\dagger\right) \biggr).
\end{align}
Now, the probability of measuring zero on the auxiliary system is given as
\begin{align}
    p(0)&=\trace{(\ketbra{0}{0}\otimes I){\rho_\mathrm{out}}}\\
    \nonumber
    &=\frac{1}{4}\left(\trace{\rho+U\rho \ee^{\ii\phi} + \rho U^\dagger \ee^{-\ii\phi} +U\rho U^\dagger}\right) \\
    \nonumber
    &=\frac{1}{2}\trace{\rho} + \frac{1}{2}\mathrm{Re}\left(\ee^{-\ii\phi}\trace{U\rho}\right),
        \nonumber
\end{align}
and similarly
\begin{equation}
    p(1) = \frac{1}{2}\trace{\rho} - \frac{1}{2}\mathrm{Re}\left(\ee^{-\ii\phi}\trace{U\rho}\right).
\end{equation}
Subtracting $p(1)$ from $p(0)$, and thereby effectively calculating the expectation value of Pauli-$Z$ on the auxiliary system, we obtain
\begin{equation}
    p(0)-p(1)=\mathrm{Re}\left(\ee^{-\ii\phi}\trace{U\rho}\right).
\end{equation}
As expected, the phase gate's phase $\phi$ allows for interpolating the imaginary and real parts of $\trace{U\rho}$.

\section{Considering post-measurement states}

Now, we add one more step to this so far standard derivation of the Hadamard test by considering the effect of measuring the auxiliary qubit, which is the first qubit of the $(1+n)$-qubit output state $\rho_\mathrm{out}$, in different Pauli bases. 
The resulting $n$-qubit post-measurement states are where we find hidden gems since we can obtain classical descriptions thereof using classical shadows.
Note that we forgo normalization here, so these are not density matrices.

Tracing out the auxiliary qubit results in
\begin{align}
\rho(I) =& \partialtrace{\mathrm{aux}}{I \rho_{\mathrm{out}}} \\
\nonumber
=& \left(\bra{0} \otimes I^{\otimes n}\right) \rho_{\mathrm{out}} \left( \ket{0} \otimes I^{\otimes n} \right)\\
\nonumber
&+ \left(\bra{1} \otimes I^{\otimes n}\right) \rho_{\mathrm{out}} \left( \ket{1} \otimes I^{\otimes n} \right)\nonumber \\
\nonumber
=&\frac{1}{4}\left( \rho  + U\rho \ee^{\ii\phi} + \rho U^\dagger \ee^{-\ii\phi}+ U\rho U^\dagger \right) \\
\nonumber
&+  \frac{1}{4}\left( \rho - U\rho \ee^{\ii\phi}-  \rho U^\dagger \ee^{-\ii\phi} +U\rho U^\dagger \right)\nonumber\\
\nonumber
=& \frac{1}{2} \left( \rho+ U\rho U^\dagger\right).
\nonumber
\end{align}
We can obtain another post-measurement state from a computational basis (Pauli-$Z$) measurement on the first qubit, given by
\begin{align}
\rho (Z) =& \partialtrace{\mathrm{aux}}{Z \otimes I^{\otimes n} \rho_{\mathrm{out}}} \\
\nonumber
=& \left(\bra{0} \otimes I^{\otimes n}\right) \rho_{\mathrm{out}} \left( \ket{0} \otimes I^{\otimes n} \right)\\
\nonumber
&- \left(\bra{1} \otimes I^{\otimes n}\right) \rho_{\mathrm{out}} \left( \ket{1} \otimes I^{\otimes n} \right)\nonumber \\
\nonumber
=&\frac{1}{4}\left( \rho  + U\rho \ee^{\ii\phi} + \rho U^\dagger \ee^{-\ii\phi}+ U\rho U^\dagger \right) \\
&-  \frac{1}{4}\left( \rho - U\rho \ee^{\ii\phi}-  \rho U^\dagger \ee^{-\ii\phi} +U\rho U^\dagger \right)\nonumber\\
\nonumber
=& \frac{1}{2} \left(  U\rho \ee^{\ii\phi} + \rho U^\dagger \ee^{-\ii\phi}\right).
\nonumber
\end{align}
While the two post-measurement states discussed above can be obtained without modification of the standard Hadamard test of Figure~\ref{fig:vanilla_hadamard}, we can also choose another measurement basis to get different post-measurement states. 
Moving into the Pauli-$X$ basis (adding a Hadamard gate and thus essentially deleting the second Hadamard gate from Figure~\ref{fig:vanilla_hadamard}), we obtain the reduced density matrix
\begin{align}
\rho (X) =& \partialtrace{\mathrm{aux}}{X \rho_{\mathrm{out}}} \\
\nonumber
= &\left(\bra{0} \otimes I^{\otimes n}\right) \rho_{\mathrm{out}} \left( \ket{1} \otimes I^{\otimes n} \right) \\
\nonumber
&+ \left(\bra{1} \otimes I^{\otimes n}\right) \rho_{\mathrm{out}} \left( \ket{0} \otimes I^{\otimes n} \right)\nonumber \\
\nonumber
=& \frac{1}{4}\left( \rho  + U\rho \ee^{\ii\phi} - \rho U^\dagger \ee^{-\ii\phi}- U\rho U^\dagger \right) \\
\nonumber
&+  \frac{1}{4}\left( \rho - U\rho \ee^{\ii\phi} +  \rho U^\dagger \ee^{-\ii\phi} -U\rho U^\dagger \right)\nonumber \\
\nonumber
=& \frac{1}{2} \left( \rho - U \rho U^\dagger \right).
\nonumber
\end{align}
Finally, we could also measure the first qubit in the Pauli-$Y$ basis and obtain
\begin{align}
\rho (Y) =& \partialtrace{\mathrm{aux}}{ Y \otimes I^{\otimes n} \rho_{\mathrm{out}}} \\
\nonumber
= &-\ii \left(\bra{0} \otimes I^{\otimes n}\right) \rho_{\mathrm{out}} \left( \ket{1} \otimes I^{\otimes n} \right) \\
\nonumber
&+ \ii \left(\bra{1} \otimes I^{\otimes n}\right) \rho_{\mathrm{out}} \left( \ket{0} \otimes I^{\otimes n} \right)\nonumber \\
\nonumber
=& -\frac{\ii}{4}\left( \rho  + U\rho \ee^{\ii\phi} - \rho U^\dagger \ee^{-\ii\phi}- U\rho U^\dagger \right) \\
\nonumber
&+  \frac{\ii}{4}\left( \rho - U\rho \ee^{\ii\phi} +  \rho U^\dagger \ee^{-\ii\phi} -U\rho U^\dagger \right)\nonumber\\
\nonumber
=& - \frac{\ii}{2} \left( U \rho \ee^{\ii \phi} - \rho U^\dagger \ee^{-\ii \phi} \right).
\nonumber
\end{align}

\section{Measurements on the system register}
\label{app:system_register}
As discussed in Refs.~\cite{faehrmannRandomizingMultiproductFormulas2022,wangQubitEfficientRandomizedQuantum2024}, measurements of the system register can lead to new quantum algorithms.
To complete our discussion of the Hadamard test, we further consider some $n$-qubit observable $O$ that we measure on the system register.
Since the trace factors for tensor products and $\trace{A}=\partialtrace{a}{\partialtrace{b}{A}}=\partialtrace{b}{\partialtrace{a}{A}}$, we find that for a single-qubit Pauli operator $P$ and $n$-qubit observable $O$,
\begin{equation}
    \trace{(P_\mathrm{aux}\otimes O_\mathrm{sys})\rho_\mathrm{out}}=\trace{O\rho(P)}.
\end{equation}
Thus, we can either estimate the expectation value of $O$ with respect to $\rho(P)$ or use random local/global Clifford measurements to obtain classical shadows of $\rho(P)$, which, given the linearity of shadows, also allows the construction 
of linear combinations of different post-measurement states.

\section{Summary of quantities accessible with the Hadamard test}
\label{app:summary}
To summarize, with the Hadamard test applied to an initial state $\rho$, we can obtain the post-measurement (observable) states and observables
\begin{align}
\rho (I) =& \partialtrace{\mathrm{aux}}{I\rho_\mathrm{out}}=\frac{1}{2} \left( \rho + U \rho U^\dagger \right)  ,\\
\braket{I\otimes O}_{\rho_\mathrm{out}}=&\trace{O\rho(I)}=\frac{1}{2}\trace{O(\rho+U\rho U^\dagger)},\\
\braket{I \otimes I^{\otimes n} }_{\rho_{\mathrm{out}}}=&\mathrm{tr} \left( \rho (I) \right) =\trace{\rho}, \\ 
\nonumber\\
\rho (X) =& \partialtrace{\mathrm{aux}}{X_\mathrm{aux}\rho_\mathrm{out}}=\frac{1}{2} \left( \rho - U \rho U^\dagger \right) ,\\
\braket{X\otimes O}_{\rho_\mathrm{out}}=&\trace{O\rho(X)}=\frac{1}{2}\trace{O(\rho-U\rho U^\dagger)},\\
\braket{X \otimes I^{\otimes n}}_{\rho_{\mathrm{out}}}=&\mathrm{tr} \left( \rho (X) \right) =0 ,\\
\nonumber\\
\rho (Y) =&\partialtrace{\mathrm{aux}}{Y_\mathrm{aux}\rho_\mathrm{out}} \\
=&-\frac{\mathrm{i}}{2} \left( U \rho \ee^{\ii \phi} - \rho U^\dagger \ee^{-\ii \phi} \right) \nonumber,\\
\braket{Y\otimes O}_{\rho_\mathrm{out}}=&\trace{O\rho(Y)}=\mathrm{Im} \left( \mathrm{tr} \left(O \ee^{\ii \phi} U \rho \right) \right)\label{eq:Y_and_O} ,\\
\braket{Y \otimes I^{\otimes n}}_{\rho_{\mathrm{out}}}
=& \mathrm{tr} \left( \rho (Y) \right) = \mathrm{Im} \left( \mathrm{tr} \left( \ee^{\ii \phi} U \rho \right) \right)\label{eq:Y_measurement},\\
\nonumber\\
\rho (Z) =& \partialtrace{\mathrm{aux}}{Z_\mathrm{aux}\rho_\mathrm{out}}\\
=& \frac{1}{2} \left( \left(  U\rho \ee^{\ii\phi} + \rho U^\dagger \ee^{-\ii\phi}\right) \right)\nonumber, \\
\braket{Z\otimes O}_{\rho_\mathrm{out}}=&\trace{O\rho(Z)}= \mathrm{Re} \left( \mathrm{tr} \left(O \ee^{\ii \phi} U \rho \right) \right),\\
\braket{Z \otimes I^{\otimes n} }_{\rho_{\mathrm{out}}} =& \mathrm{tr}(\rho (Z))= \mathrm{Re} \left( \mathrm{tr} \left( \ee^{\ii \phi} U \rho \right) \right).\label{eq:Z_measurement}
\end{align}

\section{Summary of quantities accessible with the extended Hadamard test}
\label{app:summary_extended}
Similarly, for the extended Hadamard test, including the anti-controlled application of the unitary $W$, where the output state is given by

\begin{align}
    {\rho_\mathrm{out}} = \frac{1}{4}\biggl(&\ketbra{0}{0}\otimes (W + U\ee^{\ii\phi})\rho(W + U\ee^{\ii\phi})^\dagger\\ \nonumber
    +&\ketbra{0}{1}\otimes (W + U\ee^{\ii\phi})\rho(W - U\ee^{\ii\phi})^\dagger\\ \nonumber
    +&\ketbra{1}{0}\otimes (W - U\ee^{\ii\phi})\rho(W + U\ee^{\ii\phi})^\dagger\\ \nonumber
    +&\ketbra{1}{1}\otimes (W - U\ee^{\ii\phi})\rho(W - U\ee^{\ii\phi})^\dagger \biggr),
\end{align}
we obtain the post-measurement (observable) states and observables
\begin{align}
\rho (I) =& \partialtrace{\mathrm{aux}}{I\rho_\mathrm{out}}=\frac{1}{2} \left( W\rho W^\dagger + U \rho U^\dagger \right), \\
\braket{I\otimes O}_{\rho_\mathrm{out}}=&\trace{O\rho(I)}=\frac{1}{2}\trace{O(W\rho W^\dagger+U\rho U^\dagger)}
,\\
\braket{I \otimes I^{\otimes n} }_{\rho_{\mathrm{out}}}=&\mathrm{tr} \left( \rho (I) \right) =\trace{\rho} ,
\\ 
\nonumber\\
\rho (X) =& \partialtrace{\mathrm{aux}}{X_\mathrm{aux}\rho_\mathrm{out}}=\frac{1}{2} \left( W\rho W^\dagger - U \rho U^\dagger \right) ,\\
\braket{X\otimes O}_{\rho_\mathrm{out}}=&\trace{O\rho(X)}=\frac{1}{2}\trace{O(W\rho W^\dagger-U\rho U^\dagger)},\\
\braket{X \otimes I^{\otimes n}}_{\rho_{\mathrm{out}}}=&\mathrm{tr} \left( \rho (X) \right) =0 ,\\
\nonumber\\
\rho (Y) =&\partialtrace{\mathrm{aux}}{Y_\mathrm{aux}\rho_\mathrm{out}} \\
=&- \frac{\ii}{2} \left( U \rho W^\dagger \ee^{\ii \phi} - W\rho U^\dagger \ee^{-\ii \phi} 
\right) \nonumber,\\
\braket{Y\otimes O}_{\rho_\mathrm{out}}=&\trace{O\rho(Y)}=\mathrm{Im} \left( \mathrm{tr} \left(W^\dagger O \ee^{\ii \phi} U \rho \right) \right) ,\\
\braket{Y \otimes I^{\otimes n}}_{\rho_{\mathrm{out}}}
=& \mathrm{tr} \left( \rho (Y) \right) = \mathrm{Im} \left( \mathrm{tr} \left( \ee^{\ii \phi} W^\dagger U \rho \right) \right),\\
\nonumber\\
\rho (Z) =& \partialtrace{\mathrm{aux}}{Z_\mathrm{aux}\rho_\mathrm{out}}\\
=& \frac{1}{2} \left( \left(  U\rho W^\dagger \ee^{\ii\phi} + W\rho U^\dagger \ee^{-\ii\phi}\right) \right)\nonumber ,\\
\braket{Z\otimes O}_{\rho_\mathrm{out}}=&\trace{O\rho(Z)}= \mathrm{Re} \left( \mathrm{tr} \left(W^\dagger O \ee^{\ii \phi} U \rho \right) \right),\\
\braket{Z \otimes I^{\otimes n} }_{\rho_{\mathrm{out}}} =& \mathrm{tr}(\rho (Z))= \mathrm{Re} \left( \mathrm{tr} \left( \ee^{\ii \phi} W^\dagger U \rho \right) \right).
\end{align}
For these modifications, it can also be beneficial to consider randomized approaches where $U$ and $W$ are both drawn from the same ensemble $\{p_i,U_i\}$, which produces some interesting linear combination $\sum p_i U_i$ in expectation, such as approximations of the time evolution operator~\cite{faehrmannRandomizingMultiproductFormulas2022}.\\

\section{Rigorous performance guarantees for classical shadows}
\label{app:shadow_complexity}
Since our proposed modification of the Hadamard test employs classical shadows out of the box, their well-established, rigorous performance guarantees apply here as well and readily translate into rigorous performance guarantees of our new schemes.
In detail, as rigorously presented in Refs.~\cite{huangPredictingManyProperties2020c} and further discussed in Ref.~\cite{elbenRandomizedMeasurementToolbox2023}, classical shadows are a sampling strategy with an overall $\mathcal{O}(\epsilon^{-2})$ scaling that is made more precise by the following Theorem.

\begin{theorem}[Performance guarantees for classical shadows~\cite{huangPredictingManyProperties2020c}]
    Classical shadows of size $N$ suffice to predict $M$ arbitrary linear target functions $\trace{O_1\rho},\ldots,\trace{O_M\rho}$ up to additive error $\epsilon$ given that
    \begin{equation}
        N\geq\mathcal{O}\left(\max(\norm{O_i}^2_\mathrm{shadow})\frac{\log(2M/\delta)}{\epsilon^2}\right),
    \end{equation}
    with probability at least $1-\delta$.
    The definition of the norm $\norm{O_i}_\mathrm{shadow}$ depends on the ensemble of unitary transformations used to create the classical shadow.
\end{theorem}

In practice, this means that we need to run the Hadamard test at least $N$ times, every time with a different random local or global shadow of the system register with $N$ specified as above, to obtain an error of at most $\epsilon$.

For local shadows with random Pauli measurements, the shadow norm scales as $\norm{O_i}_\mathrm{shadow}^2=\mathcal{O}(3^{w_i})$, where $w_i$ denotes the Pauli weight or locality of the Pauli observable $O_i$.
Note, however, that when using local shadows to estimate the sum of observables $\sum_i\alpha_i\trace{O_i\rho}$ up to an error of at most $\epsilon$, as would be the case for energy estimation of $\trace{H\rho}$ with $H=\sum_i\alpha_iO_i$, we need to increase the precision of the classical shadows to $\epsilon/\sum_i\abs{\alpha_i}$, resulting in more required samples.

For global shadows with random Clifford measurements, the shadow norm scales as $\norm{O_i}_\mathrm{shadow}^2=\mathcal{O}(\trace{O_i^2})$, which is closely related to the Hilbert-Schmidt norm of the observables. 
As such, they are especially powerful for observables of the form $\ketbra{\psi}{\psi}$ and useful for fidelity estimation, as also done within the Hadamard test.
The recent breakthrough of Ref.~\cite{schusterRandomUnitariesExtremely2024a} allows for random Clifford measurements with shallow quantum circuits, rendering global shadows an inexpensive addition to the Hadamard test.

Finally, we would like to note that shallow shadows as outlined in Ref.~\cite{bertoniShallowShadowsExpectation2024a} allow for depth-modulated shadows especially suited for observables that can be written as a polynomial bond dimension matrix product operator and thus allow for simultaneous estimation of local and global properties.

\end{document}